\documentclass[%
    reprint,
    aps,
    superscriptaddress,
    showpacs,
    prb,
    floatfix,
    bibnotes,
    amssymb,
    amsfonts,
]{revtex4-2}

\usepackage[utf8]{inputenc}
\usepackage[T1]{fontenc}
\usepackage{amsmath,amsthm,mathtools,mathrsfs,physics}
\usepackage{graphicx}
\usepackage{subfig}
\usepackage{hyperref}
\usepackage{color}
\usepackage{siunitx}
\usepackage{booktabs}
\usepackage{multirow}
\graphicspath{{img/}}
\usepackage{xcolor}
\definecolor{darkgreen}{rgb}{0.0, 0.5, 0.0}
\DeclareSIUnit\angstrom{\text{Å}}
\DeclareSIUnit\rydberg{Ry}
\DeclareUnicodeCharacter{2212}{-}

\newcommand{\cosns}{$\mathrm{Co}_3\mathrm{Sn}_2\mathrm{S}_2$}

\usepackage{braket}

\begin{document}
\title{Suppression of magnetism in Co$_3$Sn$_2$S$_2$ under external pressure}

\author{A. Chmeruk}
\affiliation{Theoretische Physik III, Center for Electronic Correlations and Magnetism, Institute of Physics, University of Augsburg, 86135 Augsburg, Germany}
\affiliation{Augsburg Center for Innovative Technologies (ACIT), University of Augsburg, 86135 Augsburg, Germany}
\author{D. Jones}
\affiliation{Theoretische Physik III, Center for Electronic Correlations and Magnetism, Institute of Physics, University of Augsburg, 86135 Augsburg, Germany}
\affiliation{Augsburg Center for Innovative Technologies (ACIT), University of Augsburg, 86135 Augsburg, Germany}
\author{R. Balducci}
\affiliation{Theoretische Physik III, Center for Electronic Correlations and Magnetism, Institute of Physics, University of Augsburg, 86135 Augsburg, Germany}
\author{J. Ebad-Allah}
\affiliation{Experimentalphysik II, Institute of Physics, University of Augsburg, 86159 Augsburg, Germany}
\affiliation{Department of Physics, Tanta University, 31527 Tanta, Egypt}
\author{F. Beiu\c seanu} 
\affiliation{Faculty of Science, University of Oradea, 410087 Oradea, Romania}
\author{F.~Schilberth}
\affiliation{Experimentalphysik V, Center for Electronic Correlations and Magnetism, Institute for Physics,
University of Augsburg, D-86135 Augsburg, Germany}
\author{M. A. Kassem}   
\affiliation{Department of Materials Science and Engineering, Kyoto University, Kyoto 606-8501, Japan}
\affiliation{Department of Physics, Faculty of Science, Assiut University, 71516 Assiut, Egypt}
\author{U. Schade}
\affiliation{Department of Chemistry, Humboldt-Universität zu Berlin, Brook-Taylor-Straße 2, 12489 Berlin, Germany}
\author{A. Veber}     
\affiliation{Department of Chemistry, Humboldt-Universität zu Berlin, Brook-Taylor-Straße 2, 12489 Berlin, Germany}
\affiliation{Institute for Electronic Structure Dynamics, Helmholtz-Zentrum Berlin für Materialien und Energie GmbH, Albert-Einstein-Straße 15, 12489 Berlin, Germany}
\author{L. Puskar}   
\affiliation{Institute for Electronic Structure Dynamics, Helmholtz-Zentrum Berlin für Materialien und Energie GmbH, Albert-Einstein-Straße 15, 12489 Berlin, Germany}
\author{Y. Tabata}   
\affiliation{Department of Materials Science and Engineering, Kyoto University, Kyoto 606-8501, Japan}
\author{T. Waki}   
\affiliation{Department of Materials Science and Engineering, Kyoto University, Kyoto 606-8501, Japan}
\author{H. Nakamura}  
\affiliation{Department of Materials Science and Engineering, Kyoto University, Kyoto 606-8501, Japan}
\author{C. A. Kuntscher}
\affiliation{Experimentalphysik II, Institute of Physics, University of Augsburg, 86159 Augsburg, Germany}
\author{A. \"Ostlin}
\affiliation{Augsburg Center for Innovative Technologies (ACIT), University of Augsburg, 86135 Augsburg, Germany}
\author{L. Chioncel}
\affiliation{Theoretische Physik III, Center for Electronic Correlations and Magnetism, Institute of Physics, University of Augsburg, 86135 Augsburg, Germany}
\affiliation{Augsburg Center for Innovative Technologies (ACIT), University of Augsburg, 86135 Augsburg, Germany}

\date{\today}

\begin{abstract}
The ability to control the magnetic state provides a powerful means to tune the underlying band topology, enabling transitions between distinct electronic phases and the emergence of novel quantum phenomena. In this work, we address the evolution of ferromagnetic state upon applying external pressures up to 10.8~GPa using a combined experimental and theoretical study. The standard \emph{ab initio} Density Functional Theory computation including ionic relaxations grossly overestimates the unit cell magnetization as a function of pressure. 
In our theoretical analysis we identify two possible mechanisms to remedy this shortcoming. Matching the experimental observations is achieved by a symmetry-preserving adjustment of the sulfur atoms position within the unit cell. Alternatively, we explore various combinations of the exchange and correlation parts of the effective potential which reproduce the experimental magnetization, the structural parameters and the measured optical conductivity spectra. Thus, the pressure-dependent behavior of magnetization demands a careful theoretical treatment and analysis of theoretical and experimental data.
\end{abstract}

\maketitle

\section{Introduction}
 Magnetic materials with non-trivial band topology unify many modern concepts in the theory of solids and are expected to enlarge the platform of emergent phases of matter. In these materials the intrinsic coupling between topology and magnetism enables unprecedented control over electronic states and collective excitations. Such systems present a fertile ground for realizing tunable topological transitions~\cite{hasan2010, armitage_2018}.

One such prominent magnetic material is Co$_3$Sn$_2$S$_2$ which has been identified as a ferromagnetic Weyl semimetal (WSM) with a Curie temperature of $T_c \approx 177$~K~\cite{sc.le.13,de.ho.08} and local magnetic moments of $0.33~\mu_B$/Co aligned along the crystallographic $c$-direction~\cite{ku.ok.06,sc.le.13,we.ya.15, sa.ta.15}.  
Its computed electronic structure is characterized by a gapped minority-spin channel and a Fermi level located within the majority-spin band predicting a half-metallic ferromagnetic (FM) ground state~\cite{we.an.06,sc.le.13,we.ya.15, sa.ta.15}. The states at the Fermi level are primarily originated from Co-3d orbitals. The symmetry-dependent band topology has been thoroughly discussed in the literature~\cite{xu.li.18, mo.ba.19, wa.xu.18, okamura2020_AHE, xu.zh.20, yi.zh.19, li.su.18, ro.iv.21, weyl_sign_co3sn2s2, gu.ve.20, mu.li.20}. Particularly important features of the band structure of Co$_3$Sn$_2$S$_2$ are special band crossings close to the Fermi level. Their description and understanding benefited from \emph{ab initio} modeling and various tight-binding parameterizations~\cite{de.ho.08,mo.ba.19,wa.xu.18,yi.zh.19,xu.li.18}. The exchange-split bands form two sets of Weyl loops, which merge into a four-fold Dirac loop above the Curie temperature~\cite{weyl_sign_co3sn2s2}. The presence of Weyl crossing points in the Brillouin zone (BZ), along with the observed relationship between bulk and surface states, has provided increasing evidence of potential crystal structure imperfections and their influence on the material’s physical properties~\cite{li.su.18,ya.ik.21}. The effect of electronic correlations on the stability of the WSM state was also explored by means of DFT+DMFT calculations. It was concluded that \cosns~is a moderately correlated metallic system~\cite{xu.zh.20}.

Band crossings located close to the Fermi level manifest themselves in transport properties: for example, the Berry curvature diverges in the vicinity of the Weyl nodes and thus leads to enhanced anomalous Hall conductivity~(AHC)~\cite{sh.ze.19, na.pa.20}. Interestingly, the dominant contribution to the AHC in \cosns~originates from the remnants of the gapped nodal line, as the Weyl nodes lie slightly above the Fermi level.~\cite{sh.ze.19, na.pa.20,sc.ji.22}. An alternative explanation to the enhancement of conductivity was attributed to imperfect stoichiometry leading to doping effects~\cite{li.su.18,ya.ik.21}. Controlling the positions of such as band (anti-) crossings of different dimensions (nodal points, lines, etc.) by some external parameters has become an active area of study. For example, applying an external magnetic field could lead to band reconstruction, and therefore the band crossings move to different locations in the BZ~\cite{felix2024generation}. In a similar fashion, external pressure provides an opportunity to control these topological features, influencing both their positions in momentum space and their energies. Shifting such crossings closer to the Fermi level would immediately have an effect on the various observable quantities such as optical conductivity.

A systematic experimental high-pressure study of magnetic properties of the Co$_3$Sn$_2$S$_2$ Weyl semi-metal using synchrotron X-ray diffraction (XRD), magnetization, electrical, and Hall transports has been recently published~\cite{ch.wa.19}. 
The vanishing FM order parameter at around 22 GPa was inferred by measuring AHC at different pressures. The disappearance of AHC signals the recovery of the time-reversal (TR) symmetry, meaning the transition from the FM to the paramagnetic (PM) state. This study was also supported by DFT calculations, which provided the Berry curvature distribution as well as the position of Weyl nodes under varying pressure. By contrast, another study investigated the evolution of the magnetic state up to 40~GPa and found that the AHC signal persists throughout the entire pressure range~\cite{zeng2022pressure}. A separate investigation conducted $\mu$SR experiments at various temperatures and pressures up to 2~GPa~\cite{gu.ve.20}. They report the appearance of a mixture of FM and 120$^\circ$ AFM phases close to the Curie temperature. Increasing pressure leads to a greater contribution of the AFM phase. Early studies of kagom\'e magnetism in \cosns~used mapping to models based on isotropic nearest-neighbor exchange couplings~\cite{hast.00,ya.hu.11}. Recent DFT studies, in conjunction with inelastic neutron scattering experiments, considered further-neighbor exchange interactions to account for the anisotropic magnon dispersions and linewidths observed below $T_c$~\cite{zh.ok.21, Co3sn2s2_exch_2022}. A unified intepretation of neutron diffraction and magneto-optical data suggests an easy axis ferromagnet ground state for \cosns~\cite{soh.yi.22, lee.vir.22}.

In this manuscript we discuss the impact of hydrostatic external pressure on the electronic structure of \cosns, in a complementary experimental and theoretical study of the optical conductivity. By comparing \textit{ab-initio} calculation with reported experimental data, we find that DFT functionals overestimate magnetic ordering, a tendency well known in the literature~\cite{mo.ma.93, Gunnarsson_1976}. Consequently, computations of other physical properties such as optical conductivities performed at the DFT level are expected to deviate when compared to the experimental data. In the present case, these shortcomings manifest themselves in an incorrect magnetization under increasing pressure; the standard \emph{ab initio} DFT including the ionic relaxation grossly overestimates the critical pressure at which the FM state disappears. Nevertheless, we propose two possible mechanisms within the DFT framework that correct this evidently incorrect behavior. In particular, we identify that the symmetry-preserving shift of the S atom closer to the Co-Kagomé plane seems to correct this tendency. Additionally, we investigate a possibility of employing the exchange-correlation (XC) functional with different weights of the exchange and correlation parts. We find that for certain combinations it is possible to recover the experimentally observed behavior of the magnetization as a function of pressure while also retaining reasonable lattice parameters. Among the two approaches considered, the latter provides better agreement with the experimental optical conductivities. 

The paper is structured as follows. In Sec.~\ref{sec:methods} we give a description of the methods of DFT and some of the computational details regarding the two ways of fixing the magnetization under pressure. 
In Section Sec.~\ref{sec:fm_state} we discuss the pressure induced changes in the crystal structure parameters, the unit cell magnetization and conductivities at DFT level within the generalized gradient approximation (GGA). First, we analyze the standard \emph{ab initio} DFT with no additional constraints, which produces the $M(P)$ dependence that deviates significantly from the experimental data. Next, we suggest two possible ways of fixing this issue, namely adjusting the Wyckoff position of the sulfur atom to fit the magnetization and using different proportions of exchange and correlation contributions in the effective Kohn-Sham potential. These two ways provide rather different band structures and optical conductivities. Additionally, we also study the high-temperature phase by comparing the optical conductivities with the experimental ones. Finally, in Sec.\ref{sec:summary} we draw a brief conclusion concerning our results.

\section{Computational methods and experimental details}
\label{sec:methods}
Single-crystalline \cosns{} has a rhombohedral shandite-type structure~\cite{sc.le.13} (space group $R\overline{3}m$). The conventional unit cell consists of Co layers, the so-called ``Kagomé''-layers stacked along the c axis. In a Kagomé layer, Sn atoms are surrounded by six Co atoms in a hexagonal-planar configuration ~\cite{we.an.06,de.ho.08,sc.le.13,pielnhofer_half_2014}. Another distinct Sn position exists outside the Kagomé layers which form a triangular network, visible in Fig. (\ref{fig:structure}).  Sulfur atoms form also triangular network layers~\cite{we.an.06}. The Wyckoff position of cobalt and sulfur atoms are Co-$9e (1/2, 0, 0)$ and S-$6c (0, 0, z_{S})$ respectively while the two distinct Sn atoms occupy the positions Sn1-$3b (0, 0, 1/2)$ and Sn2-$3a (0, 0, 0)$. 

The electronic structure calculations were performed using Vienna ab-initio simulation package (VASP)~\cite{blochl, kresse1996efficiency, kresse1996efficient}, at a series of target pressures, allowing all ionic degrees of freedom to change (atomic positions in the unit cell, the shape of the unit cell and the translation vectors). The structural relaxations were performed without spin-orbit coupling (SOC). The convergence of our structural optimizations was assumed when the forces on each atom were less than 10$^{-3}$ eV/\AA. The plane-wave kinetic energy cutoff was set to 400 eV. The projector-augmented plane-wave (PAW) potentials for Co, Sn, and S included the following valence states: 4-$s$, 3-$d$ and 3-$p$ for Co, 5-$s$, 5-$p$, and 4-$d$ for Sn, and 3-$s$ and 3-$p$ for S with the GGA-PBE exchange-correlation functional~\cite{pe.bu.96}. All calculations were performed using a 7-atom unit cell, and a $\Gamma$-centered \textbf{k}-grid of 16~$\times$~16~$\times$~16 was employed for Brillouin-zone integrations. The convergence of the self-consistent cycle was assumed when the total energy changes were less than $10^{-8}$ eV.

Having optimized the structure at the series of target pressures, we systematically corrected the $M(P)$ dependence. Our first approach consisted of gradually changing the Wyckoff position of the sulfur atom until the calculated magnetization per unit cell matched the experimentally observed one. The lattice parameters were fixed to the optimized values. We also tested the possibility of keeping the Wyckoff positions fixed and allowing the shape and size of the unit cell to change. Eventually, this lead to the same incorrect $M(P)$ dependence as in the case of fully unconstrained DFT.

As an alternative strategy, we varied the relative contributions of the exchange and correlation components within the GGA exchange-correlation potential~\cite{pe.bu.96}. Defining experimental lattice parameters (a, c) and the magnetization value $M$ as target properties we tuned fractions of exchange and correlation contributions into the effective DFT potential. A set of constant multiplicative weights for the exchange ($w_x$) and correlation ($w_c$) was used and kept close to reasonable ranges~\footnote{ Theoretical constrains and computational experience with existing semi-empirical fittings suggests that scaling multiplicative factors beyond $\pm 30\%$ may brake constrains leading to unphysical total energies, incorrect equilibrium lattice properties, including magnetic properties} known from the GGA exchange-correlation functional used in the DFT computation~\cite{pe.ru.08,ki.mc.21}. The procedure consists of a Bayesian calibration with Gaussian processes~\cite{ch.bl.20} applied to $w_x$ and $w_c$ and a small data set of target properties. In our empirical statistical density functional optimization, bounds for the weights were considered in the range $(0.7, 1.4)$. The experimental uncertainties for all data were taken $\sigma_{a,c} = 0.02$~\AA~and $\sigma_M =0.02~\mu_B$. We have tested such a procedure to reproduce the above target physical quantities considering the sets of data at P=10~GPa.

%
To calculate the optical conductivity, we constructed a tight-binding model using maximally localized Wannier functions (MLWF)~\cite{vanderbiltWF, Mostofi2014}. We found that high-quality MLWFs could be constructed using 8~$\times$~8~$\times$~8 \textbf{k}-grid in the DFT calculation. 
In total, 62 WFs were used. These include the $d$-orbitals 
of all Co sites 
as well as $s$- and $p$-orbitals on Sn 
and S 
sites. The optical conductivity was calculated on 120~$\times$~120~$\times$~120 \textbf{k}-grid using the following Kubo expression~\cite{Mostofi2014}:
\begin{align}
    \sigma_{\alpha \beta}(\omega) =& \sum_{n,m}\int d\textbf{k}
\frac{f_{nk} - f_{mk}}{\epsilon_n(\textbf{k}) - \epsilon_m(\textbf{k})} \nonumber \\
& \cdot \frac{\bra{\psi_{n,k}}\hat{v}_\alpha\ket{\psi_{m,k}}\bra{\psi_{m,k}}\hat{v}_\beta\ket{\psi_{n,k}}}{\epsilon_n(\textbf{k}) - \epsilon_m(\textbf{k}) - \omega + i\eta}, 
\end{align}
where $\psi_{n,k}$ is the $n$-th WF, $\hat{v}_\alpha$ is the $\alpha$-th component of the velocity operator, $f_{nk}$ is the occupation of the $(n,k)$-th state with the energy $\epsilon_n(\textbf{k})$ and $\omega$ is the frequency. Here the summation runs over all occupied ($n$) and unoccupied ($m$) bands that fall within the energy window of interest. All computations of the conductivity matrix elements include SOC.

Single crystals were synthesized and characterized as described in Ref.\ \cite{sc.ji.22}.
In the pressure-dependent studies, a diamond anvil cell (DAC) from Almax easyLab equipped with type IIA diamonds, which are suitable for infrared measurements, was used for pressure generation. The single crystals were polished to a thickness of 60 $\mu$m, and a small crystal with size of $\sim$ 190 $\mu$m $\times$ 180 $\mu$m, was loaded into the hole of a CuBe gasket inside the DAC. Finely ground CsI powder was used as quasi-hydrostatic pressure transmitting medium to ensure a well-defined sample-diamond interface during the experiment. The pressure in the DAC was determined {\it in situ} by the ruby luminescence method \cite{Mao.1986, Syassen.2008}.

Pressure-dependent reflectivity measurements at room temperature were performed in the energy range of 0.0248 to 2.728 eV (200 to 22000 cm$^{-1}$), with the polarization of the incoming radiation oriented parallel to the kagome plane.
For energies above 0.074 eV (600 cm $^{-1}$), the measurements were carried out in the lab of University of Augsburg using an infrared microscope (Bruker Hyperion), equipped with a 15$\times$ Cassegrain objective, coupled to a Bruker Vertex 80v FT-IR spectrometer. The reflectivity in the far-infrared energy range from 0.0248 to 0.074 eV was measured at the IRIS beamline of the synchrotron radiation source HZB-BESSY, using a home-built, evacuated infrared microscope, as described in Ref.\ \cite{Kuntscher.2014}, coupled to a Bruker Vertex 80v FT-IR spectrometer. The synchrotron radiation was focused on the sample using Schwarzschild objectives with a large working distance of about 55 mm and a magnification of 14$\times$.

The pressure-dependent reflectivity spectra at the sample-diamond interface R$_{s-d}$ in the energy range 0.0248 to 1.116 eV (200 to 9000 cm$^{-1}$), were determined according to $R_\text{s-d}(\omega) = [I_\text{s}(\omega) / I_\text{gold}(\omega)]$, where $I_\text{s}(\omega)$ is the intensity of the radiation reflected at the interface between the sample and the diamond anvil, $I_\text{gold}(\omega)$ is the intensity reflected from a piece of gold foil at the interface between the gold and the diamond anvil. As the reflectivity of gold shows a sharp drop at higher energies, the R$_\text{s-d}$ spectra in the energy range 1.24 to 2.728 eV (10000–22000 cm$^{-1}$) were calculated according to $R_\text{s-d}(\omega) = R_\text{dia} \times [I_\text{s}(\omega) / I_\text{dia}(\omega)]$, where $R_\text{dia}$ = 0.167 is the reflectivity of diamond, which was assumed to be pressure independent, and $I_\text{dia}(\omega)$ is the intensity reflected from the inner diamond-air interface of the empty DAC. 
In the frequency range from 1750 to 2650 cm$^{−1}$ the reflectivity spectra are affected by multi-phonon absorptions in the diamond anvils, which are not completely corrected by the reference measurement. This part of the spectrum was interpolated based on the Drude–Lorentz fitting.

\section{Results}
\label{sec:fm_state}
\begin{figure}[t!]
	\centering
	\includegraphics[width=0.48\textwidth]{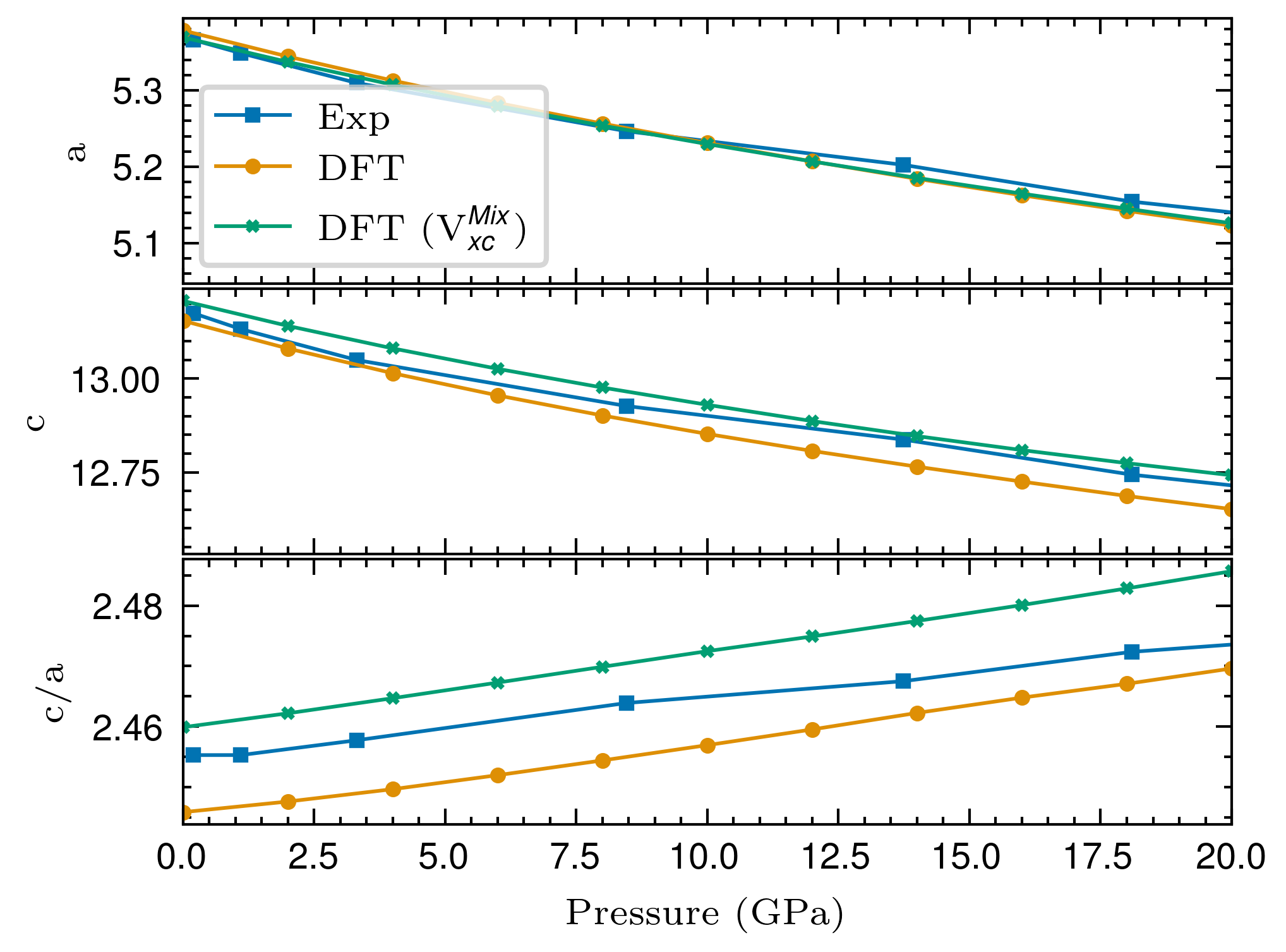}
\caption{DFT self-consistent calculation of the relaxed lattice parameters together with the experimental data of Ref.~\cite{ch.wa.19}.}
\label{fig:latpar}
\end{figure}

The standard \emph{ab initio} DFT computation of the electronic structure under pressure provides $M(P)$ dependence that significantly deviates from the experimental observation~\cite{ch.wa.19}, as seen in  Tab.~\ref{tab:lattparam}. We estimate the transition pressure from ferro- to non-magnetic (FM to NM) to be at around 51~GPa. At all pressures, the unit cell was fully relaxed, allowing all ionic degrees of freedom to change. The lattice parameters obtained in this way are shown in Fig.~\ref{fig:latpar} and are in a good agreement with experimental values. 
%
The $c$ parameter appears to be slightly underestimated in the DFT results, leading to a modest underestimation of the $c/a$ ratio. Nonetheless, both experimental and computational results follow an increasing $c/a$ trend.
Letting the atoms move during the relaxation allows us also to study possible symmetry-breaking scenarios (di-, trimerization, Jahn-Teller distortion, etc~\cite{varignon2019origin}) that might affect the $M(P)$ dependence. However, we did not observe structural changes of any kind in the entire pressure range. 

However, the only internal ionic degree of freedom that showed a systematic change was the position of the sulfur atom. We see in Tab.~\ref{tab:lattparam} that the S atom tends to move further away from the Co-Kagomé plane as the pressure increases. The S atom is positioned on the $C_{3v}$-axis, which is at the intersection of 3 mirror planes. Moving the atom along this axis leaves the point group unchanged. Hence, as our first attempt to correct the pressure dependence of magnetization, we perform such symmetry-preserving relocation of the S atom along the rotational axis. 
\begin{figure}[h]
	\centering    \includegraphics[width=0.45\textwidth]{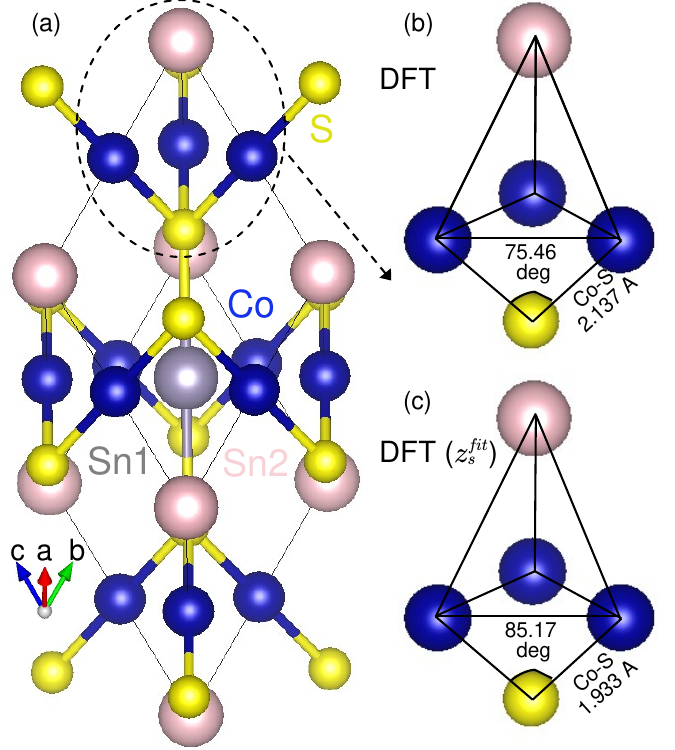}
\caption{(a) Unit cell of Co$_3$Sn$_2$S$_2$ at 10~GPa. The local environment of the Co-Co-Co triplet in the (b) fully relaxed DFT and (c) with S atom position fitted to match the observed magnetization. Relocating S atom closer to the Kagomé plane increases the Co-S-Co angle and decreases the Co-S distance, leading to a significant reconstruction of the band structure (see Fig.~\ref{fig:p10_bands})
}
\label{fig:structure}
\end{figure}

\begin{table}[b!]
    \centering
    \begin{tabular}{*{8}{c}}
        \toprule
        \multirow{2}{*}{\textbf{P(GPa)}} & 
        \multicolumn{3}{c}{\textbf{DFT}} & &
       \multicolumn{1}{c}{\textbf{DFT} (V$_{XC}^{Mix}$)} & 
       \multicolumn{1}{c}{\textbf{EXP}} \\  
        \cmidrule{2-4} 
        &  $z_S$ & $z^{fit}_S$ & M($\mu_B$) & &  M ($\mu_B$) & M$_{exp}$($\mu_B$)   \\ 
        \midrule
       0.0  & 0.28068 & 0.2600 &1.018 & & 0.906 & 1.011 \\
       2.0  & 0.28157 & 0.2590 &1.017 & & 0.849 & 0.906 \\
       4.0  & 0.28236 & 0.2570 &1.012 & & 0.786 & 0.822 \\
       6.0  & 0.28308 & 0.2575 &1.004 & & 0.719 & 0.738 \\
       8.0  & 0.28373 & 0.2585 &0.992 & & 0.652 & 0.720 \\
      10.0  & 0.28433 & 0.2620 &0.975 & & 0.587 & 0.568 \\
       \bottomrule
   \end{tabular}
   \caption{Self-consistent DFT optimized Wyckoff positions of the Sulfur atom ($z_{S}$) at different pressures. The experimental results are taken form Ref.~\cite{ch.wa.19}. 
   The 
   $z^{fit}_S$ values reproduces the experimental magnetic moments $M_{exp}$ for the relaxed unit cell parameters $(a,c)$.}
   \label{tab:lattparam}
\end{table}

This procedure allows us to study the role played by the proximity of S atoms to the Kagom\'e layers in suppressing the magnetic moment at external pressures in the framework of DFT. The constrained DFT calculations yield a fitted sulfur coordinate of $z^{fit}_S \approx 0.26$, as listed in Tab.~\ref{tab:lattparam}. With increasing pressure up to 4~GPa, $z^{fit}_S$ exhibits a slight decrease, followed by a continuous rise at higher pressures. At 10~GPa, the sulfur position slightly surpasses its ambient-pressure value ($P = 0$~GPa). In contrast, fully self-consistent DFT relaxations systematically overestimate the sulfur position by approximately 7\%, effectively placing the S atoms farther from the Kagom\'e plane defined by the Co sublattice [Fig.~\ref{fig:structure}(b, c)]. The suppression of magnetization as the S atom approaches the Kagomé plane can be understood as a consequence of enhanced overlap between the S $3p$ and Co $3d$ orbitals, which evidently weakens the effective ferromagnetic exchange between the local magnetic moments. This approach yields a sulfur Wyckoff coordinate that deviates more from the experimental value compared to fully self-consistent DFT calculations. The Rietveld refinement of the XRD data at ambient pressure gives the value of $\approx$~0.2805 for the $z$-coordinate of the S atom~\cite{expWyckoff_2023}, which is in much better agreement with the standard self-consistent DFT results. 

Semi-empirical DFT computation was performed on the data at P=10~GPa. The experimental training data combines lattice parameters and magnetization. In fact, different weights of the exchange and correlation parts lead to rather different lattice parameters during structural optimization. 
This provides a guideline for the physically relevant range of weights  when analyzing the optical conductivity. Such functional necessarily compromises between describing fundamentally different types of interactions (exchange and correlation), and the arbitrary scaling may lead to a breakdown of sum rules. We expect that the tuned weights are system specific might not fulfill certain constraints and may not be transferable as the exchange correlation model requires explicit choices for the parametrization and optimal computational performance.

The results of our Bayesian calibration are described in the following. The distribution of the exchange weight, $w_x$ is quite narrow and symmetric located roughly at 0.90. A small uncertainty of about $\pm 0.02$ was identified. The correlation weight is also well constrained at about 1.40 with slightly broader tails, the uncertainty being $\pm 0.05$. Additionally, we found  a negative statistical correlation between $w_x$ and $w_c$: when $w_x$ increases slightly, $w_c$ tends to decrease and vice versa. In other words an increasing exchange contribution can be compensated by the reduction of electronic correlations such that the total  energies varies slightly and thus, structural properties are not changed significantly. The resulting empirical-functional mix is optimal for exchange below pure DFT, and above standard correlation.
To summarize, we find that $w_x \approx 0.9 \pm 0.02$ and $w_c \approx 1.40 \pm 0.05$. The lattice constants seem to be in even better agreement than those obtained with the standard GGA functional, although the $c/a$ ratio is slightly overestimated as shown in~Fig.~\ref{fig:latpar}. The $M(P)$ dependence is also significantly improved. As for the Wyckoff position of the S atom, we find it to be very close to the standard DFT case, which is in good agreement with the experimental data (ambient pressure).

\begin{figure}[htbp]
\includegraphics[width=0.45\textwidth]{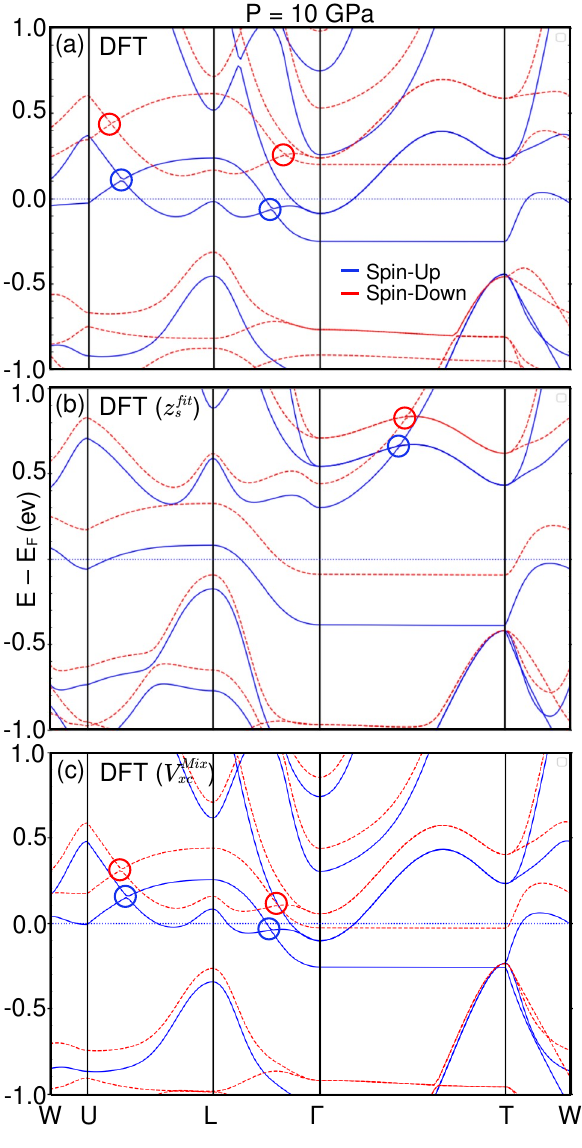}
\caption{Electronic band structure at $10$~GPa without SOC obtained with
(a) unconstrained DFT, (b) fitted Wyckoff position of the sulfur atoms and (c) with different weights of the exchange and correlation parts of the XC potential.
}
\label{fig:p10_bands}
\end{figure}

In the following, we examine the band structure and its evolution under applied pressure. As a representative case, we consider the system under an external pressure of 10~GPa. 
We analyze the electronic bands without SOC to make it simpler to track the band crossings.
    The band structures obtained with the standard GGA XC functional, by fitting the position of the S atom to the magnetization of the unit cell and with the mixed functional are shown in Fig.~\ref{fig:p10_bands} (a)-(c), respectively. The electronic bands in (a) show very little deviation from the ambient pressure reported in previous studies~\cite{xu.li.18, sc.ji.22, weyl_sign_co3sn2s2}. Near the Fermi level, we identify two pairs of band crossings associated with spin-split electronic bands. These crossings form part of a nodal line situated within the mirror plane of the Brillouin zone. As shown, this nodal line persists even at 10~GPa, consistent with the general principle that small, symmetry-preserving perturbations to the Hamiltonian may shift the position of band crossings but do not open a gap~\cite{schnyder_2008, schnyder_2016}. Since there are no non-symmorphic symmetries to protect it, the inclusion of the SOC gaps out the nodal line leaving a pair of Weyl nodes behind~\cite{fang_nodal_line_2015, fang2016topological}. The remnants of the gapped out nodal line were shown to be the source of the Berry curvature and large AHC and magneto-optical activity~\cite{sc.ji.22}. Moving on to the case where we fit the position of the S atom, Fig.~\ref{fig:p10_bands} (b), we observe a rather drastic change in the band structure. Most notably, the band crossings in the $U-L-\Gamma$ segments are no longer present. Instead, we observe the emergence of new crossings in the $\Gamma-T$ segment around 0.5 eV above the Fermi level. 
    As moving the S atom along the rotational axis does not change the point group, the nodal line is still required to exist by the present symmorphic symmetries. Therefore, we conclude that the new crossings still belong to the nodal line but was relocated to the higher energies. The system still remains metallic as we have two partially filled bands crossing the Fermi level. Thus, one expects a finite DC conductivity, but probably vanishing interband conductivity at low frequency. Lastly, we examine the band structure obtained with the mixed XC functional, Fig.~\ref{fig:p10_bands}(c). As can be seen, the shape of the bands is very similar to that obtained with the standard DFT; the spin-up nodal line is located around the Fermi level. The biggest difference is that the spin-down flat band in the $\Gamma-T$ segment has moved slightly below the Fermi level. Evidently, this is the main reason behind the reduction of the unit cell magnetization.
\begin{figure}
\includegraphics[width=0.45\textwidth]{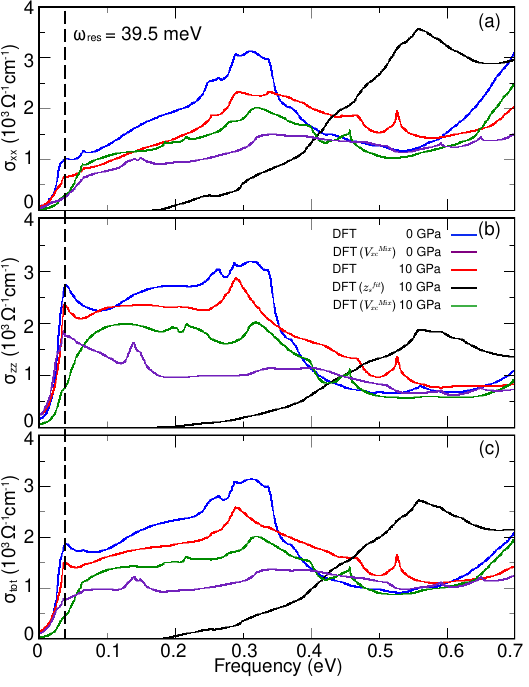}
\caption{The real parts of the computed optical conductivities at ambient and at 10~GPa:(a) the in-plane Kagom\'e $\sigma_{xx}$, (b) the out-of-plane $\sigma_{zz}$, (c) the total optical conductivity. The dashed line corresponds to the resonant frequency $(\omega_{res})$ of the nodal line~\cite{sc.ji.22}.}
\label{fig:conductivity}
\end{figure}

\begin{figure}
\includegraphics[width=0.5\textwidth]{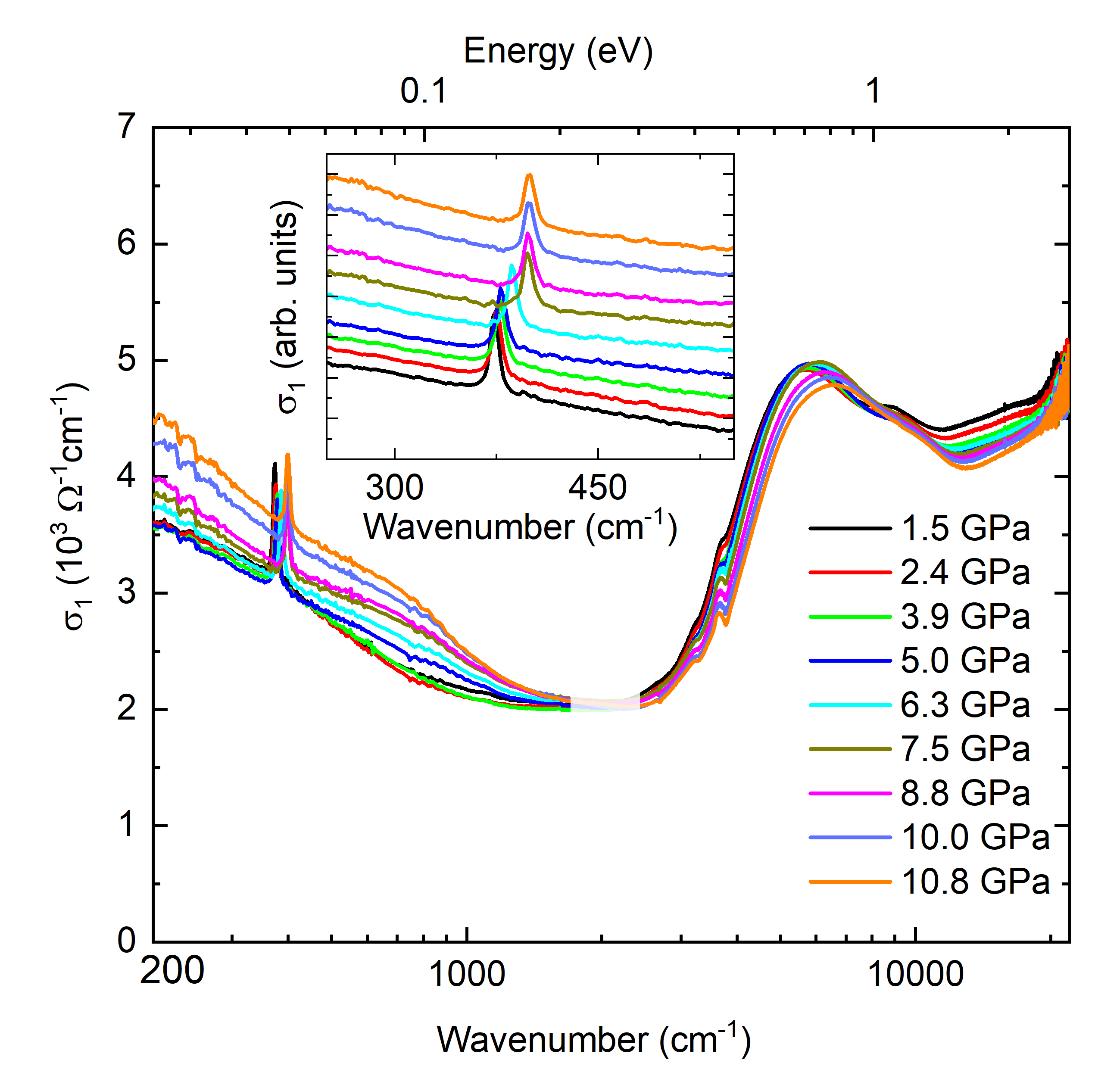}
\caption{Experimental optical conductivity of Co$_3$Sn$_2$S$_2$ single crystals in the $ab$ plane as a function of pressure. The small, pressure-independent features at around 0.45~eV are due to multiple phonon excitations in the diamond anvil, which cannot be fully corrected by the reference measurement. 
Due to the same reason, the spectra were interpolated in the range 0.22 - 0.33~eV
based on the Drude-Lorentz fit.
Inset: Phonon mode as a function of pressure (spectra are offset for clarity).}
\label{fig:exp_conductivity}
\end{figure}

Figure~\ref{fig:conductivity} shows the in-plane conductivity, $\sigma_{xx}$, in panel (a), the out-of-plane conductivity, $\sigma_{zz}$, in panel (b), and the total conductivity in panel (c), all plotted at ambient pressure (blue line), and at 10~GPa for both fully relaxed (red line) and constrained relaxed (black line) structures. At ambient pressure our computations are in excellent agreement with the previous results~\cite{sc.ji.22}. They showed that the giant anomalous AHE and the optical activity are primarily generated by the nodal line segments gapped by SOC and that Weyl points offer only a singular contribution. Another crucial factor in their interpretation is the tilting of the nodal line producing bands that remain parallel over a broad \textbf{k}-intervals and produce the so-called nodal-line resonance. Some of these interpretations can be directly applied to characterize the conductivity spectrum at 10~GPa. In particular, the conductivity remains finite at low frequencies, as the position of the nodal line remains nearly unchanged with increasing pressure. The situation changes significantly when the sulfur atom position is adjusted to match the observed magnetization. This fitting leads to substantial modifications in the band structure, with the nodal band crossings shifting away from the Fermi level towards higher energies as discussed above. 
As shown in Fig.~\ref{fig:conductivity}, the in-plane and out-of-plane conductivities $\Re{\sigma_{\alpha\alpha}}$, $\alpha=x,z$ acquire non-zero values for energies larger than 0.15$-$0.2~eV. At very low energies ($\omega \le$ 0.15~eV), the substantial reduction in interband conductivity indicates that the longitudinal components are primarily governed by transitions between the bands forming the nodal loop. According to the band structure plot Fig.~\ref{fig:p10_bands} minority as well as majority spin bands cross the Fermi level (in different sections of the BZ) possibly leading to a non-zero value of the intraband Drude conductivity. The optical conductivity calculated with $V_{xc}^{Mix}$ shows qualitatively similar features to the standard case, which is to be expected since the electronic bands are also quite similar. In particular, we still find the nodal line resonance at 39.5 meV as shown by the vertical dashed line in Fig.~\ref{fig:conductivity}, and we observe a rather strong anisotropy at low energies with the $zz$-component being larger by a factor of 2-3. 
Overall, enhancing the correlation component in the exchange-correlation functional results in reduced conductivity compared to the standard DFT. This trend is evident at both ambient pressure and at 10~GPa. A key qualitative difference between the standard and modified GGA functionals is that, in the standard case, increasing pressure tends to suppress conductivity, whereas in the modified case, it enhances conductivity. This important distinction is also evident in the experimental spectra, as shown below.

\begin{figure}
\includegraphics[width=0.47\textwidth]{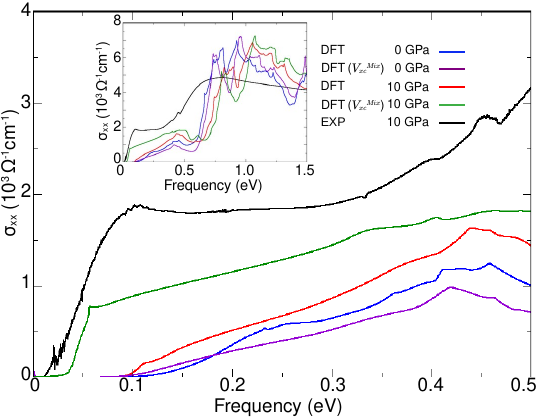}
\caption{The DFT optical conductivity calculated for NM state compared to the interband optical conductivity (polarization parallel to the kagome plane) measured at 10~GPa and room temperature (black curve). The Drude peak and the phonon mode were subtracted from the total $\sigma_1$ spectrum using a Drude-Lorentz fit. The inset shows the conductivities in the broader frequency range.} 
\label{fig:NM_conduct}
\end{figure}


To obtain the experimental optical conductivity under pressure, we carried out pressure-dependent reflectivity measurements on Co$_3$Sn$_2$S$_2$ single crystals with subsequent Kramers-Kronig analysis. 
The resulting optical condutivity is depicted in~Fig.~\ref{fig:exp_conductivity} for pressures up to 10.8~GPa. The lowest-pressure optical conductivity spectrum is in agreement with earlier reports \cite{xu.zh.20,Yang.2020,sc.ji.22,felix2024generation}, consisting of a Drude contribution due to itinerant carrier excitations and a sharp phonon mode at low energies, located around $\sim$46 ~meV. The onset of interband transitions is located at $\sim$0.01~eV, followed by two pronounced absorption bands at $\sim$0.7 and 1.1~eV and higher-energy excitations. With increasing pressure, we observe (i) an increase in the Drude term, signaling an increasing metallic character of the material, (ii) a hardening of the phonon mode (see inset of Fig.~\ref{fig:exp_conductivity}), and (iii) a blue-shift of the interband excitations. Overall, the pressure-induced changes in the optical conductivity spectrum are modest and the profile is unchanged up to the highest measured pressure (10.8~GPa). 

In order to directly compare theoretical and experimental spectra we extracted the interband optical conductivity from the experimental $\sigma_1$ spectrum, by applying a simultaneous Drude-Lorentz fit of the reflectivity and optical conductivity data and considered the pressure-dependent dc transport data of Ref.~\cite{ch.wa.19}. 
By subtracting the Drude term related to the intraband excitations and the Lorentz term ascribed to the phonon mode, we obtain the interband optical conductivity, which can be directly compared with the theoretical optical conductivity. The measured interband optical conductivity at 10~GPa can be well reproduced by the theoretical calculations. 
Although the absorption edge above 0.4~eV is significantly broadened in the experiment, likely due to lifetime effects of the electronic excitations, the low energy spectrum is qualitatively better reproduced by our modified DFT approach using the mixed XC functional.

\section{Summary and discussions}
\label{sec:summary}
We present results of a combined theoretical and experimental study on magnetic and optical properties of \cosns~ in presence of external pressures up to 10.8~GPa. 

On the theoretical side we demonstrate that the standard \emph{ab initio} DFT produces incorrect $M(P)$ dependence, significantly overestimating the critical pressure at which the absence of the FM state was experimentally reported~\cite{ch.wa.19}. 
In an attempt to understand the source of the discrepancy between the experimental and theoretical results we considered an alternative computational scenario that eventually improved the experiment/theory agreement at the cost of stepping aside from the fully \emph{ab initio} description of the electronic structure. Namely, we performed a symmetry-preserving adjustment of the sulfur atom position inside the unit cell which lead to improved values of  magnetisation at different pressures. This approach results in significant modifications of the band structures compared to the standard DFT calculations. In particular, we find that the nodal line is no longer situated near the Fermi level, but instead shifts to higher energies (approximately $\sim$0.5~eV above the Fermi level). Its existence, however, remains guaranteed, as displacing the S atom along the $C_{3v}$ rotational axis does not alter the space group. The band structure obtained using this method is different in comparison with~\cite{ch.wa.19}: the position of the nodal line (Weyl nodes when SOC is turned on) is much higher above the Fermi level, and consequently the optical conductivity shows almost no interband transitions up to $\sim$0.2~eV. In a second attempt to resolve the pressure dependence of magnetization we have also attempted to employ a semi-empirical PBE-GGA exchange-correlation functional constructed from a reduced data set with targeted properties such as the experimental lattice parameters and the magnetization. The data has been obtained by running self-consistent DFT computations with different multiplicative factors added to the exchange and correlation parts of the GGA-PBE functional. Our Bayesian calibration produced uni-modal narrow distributions for the exchange $0.9 \pm 0.02$ and $1.4\pm 0.05$ correlation multiplicative factors. Our modified DFT resolves the $M(P)$ dependence issue and also yields electronic band structures and optical conductivities that show much better agreement with previously reported data.

On the experimental side, we provide a description of the methodology of reflectivity measurements and show results of optical conductivities at room temperature for a series of pressures, which to our best knowledge were not presented in the literature. Overall, the data show that upon increasing pressure the metallic character is enhanced and interband excitations are shifted to higher energies. In addition, the measured optical conductivity spectra were compared to the DFT conductivities. Using the optimized semi-empirical functional for the non-magnetic state we also find it to produce the optical conductivity that is in better agreement with our experimental findings. We also find that the low-frequency conductivity is extremely sensitive to the Fermi level position. A shift of merely 0.01 eV upward (electron doping) or downward~(hole doping) causes a noticeable increase or decrease in this region of the optical conductivity. This pronounced dependence implies that even slight non-stoichiometry could significantly affect the results, potentially contributing to our underestimation of the optical conductivity relative to experimental measurements.

Finally, we would like to comment on a certain aspects of DFT modeling presented above. Although the interpretation of experimental results (lattice structure, magnetization, and optical conductivity) at ambient pressure is well captured within the DFT and documented in many publications, the results for P$\neq$0~GPa show certain shortcomings of the modeling. 
Results obtained using semi-empirical DFT functionals that target a limited number of materials properties has to be interpreted with caution. Within the DFT the self-interaction error is known to be the cause for the inaccurate exchange-correlation functionals~\cite{becke1993density}. Exchange tends to localize electrons while correlation tends to delocalize them. As our Bayesian calibration (performed on the P=10~GPa targeted data) provides $w_x<1$ and $w_c>1$ the optimized (empirical-)functional becomes more de-localizing, i.e. the electronic correlation tend to spread the electronic density to reduce Coulomb repulsion. The behavior of the optical conductivity presented here is therefore rather well captured by the computation. As $w_c>1$ a valid description of physical properties require stronger electronic  correlations, so methods like DFT+DMFT~\cite{ko.sa.06} maybe used to complement the correlation effects of DFT functionals. At ambient pressures DFT+DMFT computations on \cosns~have already been reported~\cite{xu.zh.20} with a reasonable success, which is expected to be extended at nonzero pressures. Additionally, to address the ferromagnetic transition under pressure requires the possibility to model the existence of local moments and their formation. While this is possible within the DFT+DMFT framework, on considerable computational cost, 
the special quasirandom structures (SQS) approach~\cite{zunger1990special, wei1990electronic} might be a possible alternative. Practically, the SQS approach requires a large unit cell, which can accommodate various local FM and AFM environments. Although some attempts along these lines a;ready exist, a complete analysis using these methods for Kagomé systems to our best knowledge remain a considerable challenge.

\begin{acknowledgments}
We thank HZB-BESSY for the allocation of beamtime at beamline IRIS.
We acknowledge technical support by R. Borkenhagen, M. K\"opf, M. Lamp, S. Rojewski during the beamtime at HZB-BESSY.
This work was supported by the Deutsche Forschungsgemeinschaft (DFG, German Research Foundation)–TRR 360–492547816.
\end{acknowledgments}

\bibliographystyle{apsrev-title.bst}
\bibliography{sources.bib}

\end{document}